\documentclass[aps,prb,amsmath,amssymb,10pt,a4paper,floatfix,twocolumn,superscriptaddress]{revtex4-1}
\usepackage[colorlinks,bookmarks=true,citecolor=blue,linkcolor=red,urlcolor=blue,citecolor=blue]{hyperref}

\usepackage{epsfig}
\usepackage{graphicx}
\usepackage{dcolumn}
\usepackage{bm}
\usepackage{bookmark}
\usepackage{color}
\usepackage{ulem}

\begin{document}
\title{Entanglement Entropy of Non-Hermitian Free Fermions}
\author{Yi-Bin Guo}
\affiliation{Beijing National Laboratory for Condensed Matter Physics and Institute of Physics, Chinese Academy of Sciences, Beijing 100190, China}
\affiliation{School of Physical Sciences, University of Chinese Academy of Sciences, Beijing 100049, China}
\author{Yi-Cong Yu} \email{ycyu@wipm.ac.cn}
\affiliation{Beijing National Laboratory for Condensed Matter Physics and Institute of Physics, Chinese Academy of Sciences, Beijing 100190, China}
\affiliation{State Key Laboratory of Magnetic Resonance and Atomic and Molecular Physics, Wuhan Institute of Physics and Mathematics, IAPMST, Chinese Academy of Sciences, Wuhan 430071, China}
\author{Rui-Zhen Huang} \email{huangrzh@icloud.com}
\affiliation{Kavli Institute for Theoretical Sciences, University of Chinese Academy of Sciences, Beijing 100190, China}
\author{Li-Ping Yang}
\affiliation{Department of Physics, Chongqing University, Chongqing 401331, China}
\author{Run-Ze Chi}
\affiliation{Beijing National Laboratory for Condensed Matter Physics and Institute of Physics, Chinese Academy of Sciences, Beijing 100190, China}
\affiliation{School of Physical Sciences, University of Chinese Academy of Sciences, Beijing 100049, China}
\author{Hai-Jun Liao}
\affiliation{Beijing National Laboratory for Condensed Matter Physics and Institute of Physics, Chinese Academy of Sciences, Beijing 100190, China}
\affiliation{Songshan Lake Materials Laboratory, Dongguan, Guangdong 523808, China}
\author{Tao Xiang}
\affiliation{Beijing National Laboratory for Condensed Matter Physics and Institute of Physics, Chinese Academy of Sciences, Beijing 100190, China}
\affiliation{School of Physical Sciences, University of Chinese Academy of Sciences, Beijing 100049, China}
\affiliation{Beijing Academy of Quantum Information Sciences, Beijing 100193, China}

\date{\today}

\begin{abstract}
We study the entanglement properties of non-Hermitian free fermionic models with translation symmetry using the correlation matrix technique.
Our results show that the entanglement entropy has a logarithmic correction to the area law in both one-dimensional and two-dimensional systems. For any one-dimensional one-band system, we prove that each Fermi point of the system contributes exactly $1/2$ to the coefficient $c$ of the logarithmic correction.
Moreover, this relation between $c$ and Fermi point is verified for more general one-dimensional and two-dimensional cases by numerical calculations and finite-size scaling analysis. In addition, we also study the single-particle and density-density correlation functions. 
\end{abstract}
\maketitle

\section{Introduction}
Entanglement plays a significant role in various fields, including quantum information, quantum gravity, statistical physics, and condensed matter physics\cite{amico2008entanglement,Eisert2010,nishioka2018entanglement}. It captures the most primary feature of quantum physics and has become a widely used concept in modern physics. For quantum systems, the bipartite entanglement entropy (EE) quantifies to what degree their quantum states are entangled. In the past decades, the applications of EE are booming in various fields in physics. For instance, in the low dimensional computational physics,  the scaling behavior of EE determines the validities of many mainstream numerical techniques, like the density renormalization group, matrix product states method, etc.; in the modern cosmology,  the holographic entanglement entropy shed new lights on the mesmerizing quantum gravity\cite{ryu2006aspects,ryu2006holographic}. Much inspiring progress also appears in understanding the EE properties of quantum phases and quantum phase transitions\cite{amico2008entanglement,Eisert2010,TopS1,TopS2,TopS3,area0,hastings2007entropy,hastings2007area,brandao2013area}. 
It has been shown that the EE of the ground state usually obeys the area law for local gapped systems\cite{amico2008entanglement,Eisert2010,area0,hastings2007entropy,hastings2007area,brandao2013area}. In particular, for a gapped topological ordered state, the EE has an extra universal constant correction, namely topological entanglement entropy, which has become the important quantity to characterize topological order\cite{amico2008entanglement,Eisert2010,nishioka2018entanglement,TopS1,TopS2,TopS3}. While for a gapless system, the EE of the ground state usually has logarithmic correction besides the area law\cite{amico2008entanglement,Eisert2010,nishioka2018entanglement,swingle2010entanglement,QCPS1,jin2004quantum,QCPS3}. Especially for a lot of one-dimensional ($1D$) quantum critical systems, their entanglement properties can be well understood via the conformal field theory (CFT), and the EE is closely related to the conformal anomaly\cite{nishioka2018entanglement,Calabrese2009EntanglementEA}.

On the other hand,  the non-Hermitian systems are receiving increasing attention from different research communities in recent years \cite{Lee2014,daley2014quantum,Nakagawa2018,li2019observation,wu2019observation,Yamamoto2019,Xiao2019,ashida2020non}. This research upsurge is driven by the breakthrough from experiment and theory. Experimentally, non-Hermitian systems already have been implemented in photonic crystals, biological systems, mechanical systems, quantum optical systems, etc\cite{ashida2020non,longhi2018parity,guo2009observation,ruter2010observation,bertoldi2017flexible,chou2011non,lebrat2019quantized,schomerus2013topologically,zhen2015spawning,T5}. Theoretically, many unique properties absent in Hermitian systems have been revealed, including non-Hermitian skin effect, modified bulk-boundary correspondence, generalized topological phases, new non-Hermitian universalities\cite{ashida2020non,xu2017weyl,YaoSkin2018,yoshida2019symmetry,OkumaSkin2020,borgnia2020non,T1,yang2020non,zhang2020correspondence,T4,T6,T7,T8,kawabata2019topological,T10,T11,T12,T13,T14,wang2020state,tianyu2021two,ren2020photon,Jin2021}, and so on. Because the EE plays the central role in the  many-body physics,  the study of the EE of the non-hermitian system has experimental significance.

There were some significant and pioneering researches focusing on the entanglement properties in the non-Hermitian systems\cite{couvreur2017entanglement,herviou2019entanglement,lee2020many,chang2020entanglement,mu2020emergent,chen2020entanglement,okuma2021quantum,modak2021eigenstate,bacsi2021dynamics}. 
However, many crucial problems yet need to be explored intensively. For example, the interplay between the EE and the Fermi surface remains elusive for the non-Hermitian fermionic systems, especially in two and higher dimensions.
Also, for the general properties of the EE in non-Hermitian systems, it is unclear to what extent they retain compared to those in the Hermitian cases. 
These questions motivate us to find the answers and summarize what quantifies the entanglement entropy in non-Hermitian free fermionic systems.

In this paper, we use the correlation matrix technique to systematically investigate the general properties of the EE in the non-Hermitian free fermionic systems. We analytically prove that the EE of $1D$ one-band systems obey the area law with a logarithmic correction and each Fermi point exactly contributes $1/2$ to the coefficient $c$ of the logarithmic correction. These properties of the EE are numerically confirmed in more complicated $1D$ multi-band systems and higher dimensional systems. These results reveal the interesting results that the entanglement entropy relies on the topology of the Fermi surface for both hermitian and non-hermitian systems.

The remaining parts are organized as follows. In Sec.~\ref{sec:method}, we introduce the definition of the EE in the non-Hermitian systems and related calculation methods. In Sec.~\ref{sec:analytic}, we present the exact results of the EE of one-band systems in one dimension. 
In Sec.~\ref{sec:numerics}, we extend our considerations to $1D$ multi-band model and two-dimensional ($2D$) systems. Finally, we summarize in Sec.~\ref{sec:summary}.

\section{Method}
\label{sec:method}

\subsection{Diagonalization of Non-Hermitian Fermionic Systems}
The Hamiltonian of a general non-Hermitian free fermionic model is expressed as
\begin{equation}
H=\sum\limits_{ij} c_{i}^{\dagger }\mathcal{H}_{ij} c_{j},
\label{Eq:H}
\end{equation}
where $\mathcal{H}_{ij}$ denotes a hopping matrix element. The asymmetry of $\mathcal{H}_{ij}$ under the exchange of $i$ and $j$ leads to the result that $\mathcal{H}_{ij}$ can not be diagonalized by a unitary transformation like Hermitian systems. However, one can still diagonalize it with a general matrix $U$, namely, 
\begin{equation}
   H=\sum_{m} {\varepsilon_m} f_{Rm}^{\dagger }f_{Lm}, 
\end{equation}
where the operators of left and the right fermionic modes are respectively
\begin{eqnarray}
f_{Lm} =\sum_{j}U^{-1}_{mj}c_{j},   \hspace{0.5cm}
f_{Rm}^{\dagger } =\sum_{i}c_{i}^{\dagger }U_{im}. 
\label{Trans}
\end{eqnarray}%
It is worth noting that the annihilation operators of the left fermionic modes in Eq.~(\ref{Trans}) are exactly the conjugate of the creation operators of the right fermionic modes of $H^{\dagger}$, which is common definition in literatures. Besides, for the model with symmetry $O\mathcal{H}=\mathcal{H}^{\dagger}O$ and $D^{\dagger}=T^{-1}DT$ with $T$ is a self-adjoint matrix (assume that the eigen spectrum of $H$ and $H^{\dagger}$ are the same, which is satisfied in the model we studied in this article) the general matrix $U$ satisfies $TU^{\dagger}O=U^{-1}$. The significant difference between a non-Hermitian system and a Hermitian system lies in that
$f_{Lm}^{\dagger }\neq (f_{Rm})^{\dagger }$ due to $U^{-1} \neq U^{\dagger}$
. However, the anti-commutation relations still hold,
\begin{equation}
\{ f_{Rn}^{\dagger },f_{Lm}\} =\delta _{mn},
\{ f_{Lm}^{\dagger },f_{Ln}^{\dagger }\} =\{f_{Rm},f_{Rn}\}=0. 
\end{equation}
Since the energy $\varepsilon_m$ in a non-Hermitian system is not necessarily real, the definition of the ground state is still controversial. One way to construct the left and right ground state is filling up the levels whose real part of the energy is less than the certain Fermi energy $\varepsilon_{F}$ (referred to the RF case)\cite{chang2020entanglement,herviou2019entanglement} 
\begin{equation}
\langle \mathrm{G_L} | = \langle 0 | \prod\limits_{\mathrm{Re}(\varepsilon
_{m}) < \varepsilon_{F}} f_{Lm},  \,
| \mathrm{G_R} \rangle =\prod\limits_{\mathrm{Re}(\varepsilon
_{m}) < \varepsilon_{F}}f_{Rm}^{\dagger} |0 \rangle,
\end{equation}
here $| 0 \rangle$ denotes the vacuum state. Similarly, to study how the geometry of the Fermi surface affect the entangelement entropy, we can also define a different ground state by filling up the levels according to the imaginary part of the spectra $\mathrm{Im}(\varepsilon_{m}) < \varepsilon_{F}$ (referred to as the IF case).

\subsection{Entanglement entropy and correlation matrix technique}
Usually, the EE of the ground state of a system can be computed by the reduced density matrix (RDM). To be concrete, the entanglement entropy
\begin{equation}
S = -\text{Tr}[\rho_{A}\log \rho_{A}],
\end{equation}
where ${\rho}_A = \text{Tr}_{\overline A}(\rho)$ is the reduced density matrix of a subsystem by tracing out the environment part $\overline A$.

For a non-Hermitian system, there are usually three types of density matrix since the left eigenstates is not conjugate to the right ones. The first definition of the reduced density matrix is the right density matrix ${\rho}^{RR}$, which is defined as
\begin{equation}
{\rho}^{RR} = \frac{| \mathrm{G_R} \rangle \langle \mathrm{G_R} |}{\text{Tr}(| \mathrm{G_R} \rangle \langle \mathrm{G_R} |)}.
\label{Eq:rhoRR}
\end{equation}
Similarly, the second definition is the left density matrix ${\rho}^{LL}$, replacing $|\mathrm{G_R} \rangle$ with $|\mathrm{G_L} \rangle$. The last definition is the biorthogonal density matrix ${\rho}^{RL}$ 
\begin{equation}
{\rho}^{RL} = \frac{| \mathrm{G_R} \rangle \langle \mathrm{G_L} |}{\text{Tr}(| \mathrm{G_R} \rangle \langle \mathrm{G_L} |)}.
\label{Eq:rhoRL}
\end{equation}
The $\rho^{RL}$ corresponds to biorthogonal quantum mechanics, which forms a self-consistent theoretical framework for non-Hermitian systems\cite{brody2013biorthogonal}. Thus, we only consider the EE calculated with the $\rho^{RL}$ in this paper.

There are usually three methods to calculate the reduced density matrix, including the singular value decomposition(SVD) method, the correlation matrix technique, and the overlap matrix method\cite{peschel2009reduced,fishman2015compression,chang2020entanglement}. Specifically, the SVD method often encounters the so-called exponential-wall problem and hence only applies to small systems. By contrast, the computational cost of the latter two methods only increases polynomially with the system size. In addition, the overlap matrix method is similar to the correlation matrix technique. In the following, we will review the correlation matrix technique. The correlation matrix is defined as\cite{chang2020entanglement,herviou2019entanglement}
\begin{equation}
C^{A} _{ij}=\langle \mathrm{G_L} | c_{i}^{\dagger
}c_{j} | \mathrm{G_R} \rangle
=\sum\limits_{m=1}^{N_{e}}U^{-1}_{mi}U_{jm},
\label{eq:CM}
\end{equation}
where $i$ and $j$ are the sites inside subsystem $A$, and $N_e$ is the number of occupied states whose energies satisfy $\mathrm{Re}(\varepsilon_{m}) < \varepsilon_{F}$ for the $\mathrm{RF}$ case ($\mathrm{Im}(\varepsilon_{m}) < \varepsilon_{F}$ for the $\mathrm{IF}$ case). Then the reduced density matrix $\rho_A$ for subsystem $A$ can be calculated directly from the above correlation matrix $C^{A}$ by the Wick theorem\cite{chang2020entanglement,herviou2019entanglement}. Then the entanglement entropy $S$ of subsystem $A$ can be obtained by \cite{QCPS3,peschel2009reduced,fishman2015compression}
\begin{align}
S = -\sum\limits_{m}\big[\zeta_m\log{\zeta_m} + (1-\zeta_m)\log(1-\zeta_m)\big], 
\label{entropy}
\end{align}
where $\zeta_m$ is the eigenvalue of the correlation matrix $C^{A}_{ij}$.
In general, the entanglement spectra of the non-Hermitian system are complex. However, for all the models studied in this work, we observe that $\zeta_m$ is real by careful scaling analysis (there is an extremely small imaginary part that gradually vanishes with the system size increases, as shown in Fig.~\ref{smImagScaling}).

\section{Exact Results of $1D$ Non-Hermitian Free Fermionic System}
\label{sec:analytic}
For one-dimensional Hermitian non-interacting gapless systems, the EE has been well understood by CFT. However, to our knowledge, there is no relevant exact result or general theory about the EE in the non-Hermitian free fermionic systems. In this section, we present the exact results for a general $1D$ non-Hermitian Hamiltonian $H$ with only one energy band and periodic boundary. $H$ here can contain general long-range hopping terms and subsequent arbitrary complicated Fermi surface structures. Generally, the Fermi surfaces can be separated into many separate parts $\mathcal{FS} = \bigcup_{i=1}^{N_f/2} [k_{2i-1},k_{2i}]$, where $N_f$ is the number of Fermi points, and the momentum of Fermi points satisfies $0<k_1<k_2<\ldots <k_{N_f} < 2\pi$.

The one-band Hamiltonian can be directly diagonalized by the unitary Fourier transformation $U_{jm}=\frac{1}{\sqrt{N}}e^{\mathrm{i}\frac{2{\pi}mj}{N}}$. Then in the thermodynamic limit, the correlation matrix shown in Eq.~(\ref{eq:CM}) can be written as
\begin{align}
C^A_{ij} = \int_{\mathrm{occ.}} \frac{\mathrm{d}p}{2\pi} \,
\mathrm{e}^{-\mathrm{i}p(i-j)},
\end{align}
in which $\mathrm{occ.}$ denotes the occupied momentum points. The eigenvalues of the correlation matrix $C^A$ are obtained by evaluating
the characteristic polynomial of $\det{(\lambda I - C^A)}$~\cite{jin2004quantum}, where 
\begin{align}
(\lambda I - C^A)_{ij} = \int_{0}^{2\pi} \frac{\mathrm{d}p}{2\pi} g(p)\mathrm{e}^{-\mathrm{i}p(i-j)}
\label{eq:C_lambda}
\end{align}
and the kernel function $g(p)$ are defined as follows~\cite{CIT-006}
\begin{align}
g(p) = \Big{\lbrace}
\begin{array}{cc}
\lambda - 1 & \quad p \in \mathrm{occ.}\vspace{1ex},   \\
\lambda & \quad \text{otherwise}  .
\end{array}
\label{gen}
\end{align}
The $g(p)$ can also be factorized to an analytic function $\psi(p)$ and $N_f$ periodic functions $t_{\beta_r,k_r}(p)$,
namely,
\begin{align}
g(p) = \psi(p) \prod_{r=1}^{N_f} t_{\beta_r,k_r}(p),
\label{factorg}
\end{align}
where $\beta_{r} =(-1)^{r}\beta_{\lambda}$ by defining
\begin{equation}
    \beta_{\lambda} = \frac{1}{2\pi \mathrm{i}}\log\left(\frac{\lambda - 1}{\lambda}\right),
\end{equation}
and $k_r$ is the momentum of $r$-th Fermi point. Besides, $\psi(p)$ is denoted as
\begin{equation}
\begin{split}
\psi(p) = \lambda{\prod_{l=1}^{N_f/2}\left(\frac{\lambda - 1}{\lambda}\right)^{\frac{k_{2l}-k_{2l-1}}{2\pi}}}
\end{split}
\end{equation}
with
\begin{equation}
t_{\beta_r,k_r}(p)
= \exp[\mathrm{i}\beta_r(\text{mod}(p-k_r, 2\pi)-\pi)].
\end{equation}
Within this formalism, the characteristic polynomial of
the correlation matrix Eq.~(\ref{eq:C_lambda}) can be obtained via the Fisher-Hartwig theorem \cite{fisher1969toeplitz,basor1979localization}
\begin{align}
D_L(\lambda)=&\prod_{i \neq j} [1-\mathrm{e}^{\mathrm{i}(k_i-k_j)}]^{\beta_i \beta_j}
[G_B(1+\beta_{\lambda})G_B(1-\beta_{\lambda})]^{N_f} \notag \\
& \left[\lambda\prod_{l=1}^{N_f/2}\left(\frac{\lambda-1}{\lambda}\right)
 ^{\frac{k_{2l}-k_{2l-1}}{2\pi}}\right]^L
  L^{-{\beta_{\lambda}}^2 N_f},
 \label{eq:DL}
 \end{align}
 where $L$ is the system size and the $G_B(z)$ is the Barnes G-function defined
 as 
 \begin{align}
 G_B(1+z)=&(2\pi)^{z/2}\exp[-(z+1)z/2-\gamma_E z^2/2] \notag\\
        &\prod_{n=1}^{\infty}(1+z/n)^n\exp[-z+z^2/(2n)]
 \label{eq:Gfunction}
 \end{align}
 with $\gamma_E$ is the Euler constant.
 Eventually, the EE can be evaluated from the
 above characteristic polynomial of the correlation matrix as~\cite{jin2004quantum,Its2006Entropy}
 \begin{align}
 S = \frac{1}{2\pi\mathrm{i}}\oint_{\mathcal{C}} W(\lambda) \mathrm{d}
 \log D_L(\lambda),
 \end{align}
 where $W(\lambda)=-\lambda \log{\lambda} - (1-\lambda)\log(1-\lambda)$ and $\mathcal{C}$ is the contour circling the interval $[0,1]$ in the real axis of the complex plane in the anticlockwise direction.
 
 The explicit calculation \cite{jin2004quantum} shows that the
 leading divergent term of $S$ increases logarithmically with $L$. This term is solely determined by the power of $L$ in Eq.~(\ref{eq:DL}).
Thus the asymptotic formula for the EE reads
\begin{align}
S = \frac{c}{3} \log(L) + O(1),
\label{eq:Exact_SA}
\end{align}
where $c=\frac{N_f}{2}$ in the thermodynamic limit $L \to \infty$.

It seems that the EE for the $1D$ multi-band or higher-dimensional free fermionic systems can also be achieved by analyzing the correlation matrix. However, it will become extremely challenging, because in those cases the correlation matrices turn out to be block-Toeplitz matrices. As a result, one can no longer use the Fisher-Hartwig theorem to analyze the EE. We take the $2$-sublattice model defined in Eq.~(\ref{Eq:H1d}) as an example to explain the above statement and one can find the details in Appendix[\ref{subsec:ExactDiag}]. The rigorous mathematical proof is still absent for the general case. Do the entanglement properties like Eq.(\ref{eq:Exact_SA}) still hold? We attempt to address this issue by numerical simulation. 

\section{Numerical Results}
\label{sec:numerics}
In this section, by the numerical simulations, we investigate the entanglement properties of the more general systems with periodic boundary conditions, including $1D$ multi-band free fermionic systems and $2D$ non-Hermitian free fermionic systems. 

\begin{figure}[t]
	\includegraphics[angle=0,scale=0.33]{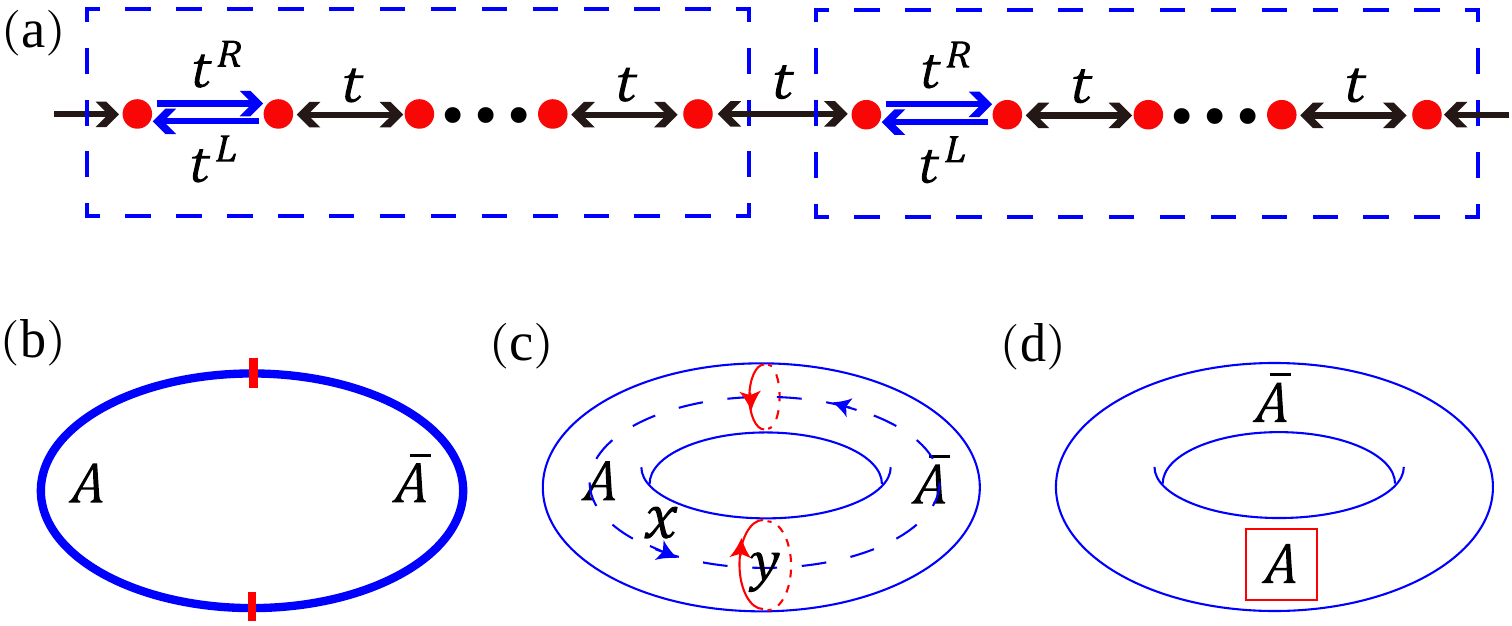}
	\caption{(a) Schematic diagram of the model in Eq.~(\ref{Eq:H1d}), where the dashed box represents a unit cell. The non-Hermitian and Hermitian bonds are colored blue and black, respectively. (b-d) show the partition boundary for the bipartite entanglement entropy for the $1D$ chain, the quasi-one-dimensional (quasi-$1D$) ladder, and the $2D$ lattice. The system $A$ and the environment $\overline A$ are partitioned by the red line.}
	\label{fig:1D_model}
\end{figure}

\subsection{$1D$ System}
Firstly we consider a $1D$ spinless fermions model with $n$-sublattice as shown in Fig.~\ref{fig:1D_model}(a),
\begin{align}
H = \sum_{i=1}^{L} (t_i^L c_{i}^{\dagger} c_{i+1}
+ t_i^R c_{i+1}^{\dagger} c_{i}),
\label{Eq:H1d}
\end{align}
with the hopping constants
\begin{align}
     \Big{\lbrace}
    \begin{array}{ll}
    &t_{i}^L = t^L,  \,\, t_{i}^R = t^R,  \,\,\quad \text{if } i =n(l-1)+1\vspace{1.2ex}, \\
    &t_{i}^L = t_{i}^R = t,  \qquad\qquad \text{otherwise}.
    \end{array}
\label{Eq:Hopping}
\end{align}
Here $l$ denotes the $l$-th unit cell, $t = 1$, $t^L = 1+\gamma/2$, and $t^R = 1-\gamma/2$. The total number of sites is $L=n N_c$, where $n$ denotes the number of sites in a unit cell, $N_c$ is the number of the unit cell. 

\subsubsection{Entanglement Entropy}
\label{subsec:EE of 1D}
\begin{figure}[t]
	\includegraphics[angle=0,scale=0.101]{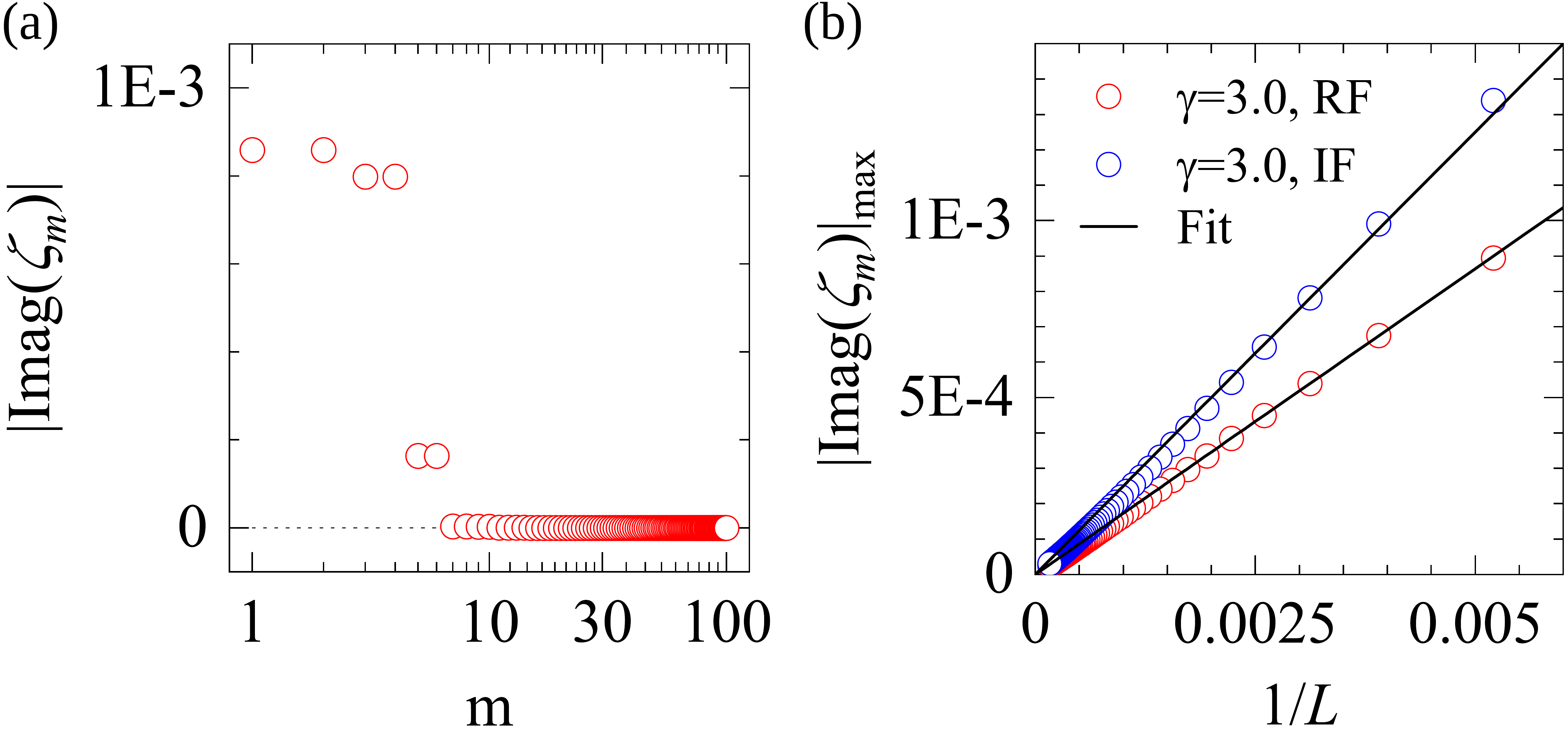}
	\caption{The imaginary part of the sub-system correlation matrix eigenspectra for the half-filling $2$-sublattice model described by Eq.~(\ref{Eq:H1d}). (a) The imaginary part of the correlation matrix eigenspectra for a $L=200$ chain. (b) The finite-size scaling of the maximal imaginary part for the correlation matrix eigenspectra.}
	\label{smImagScaling}
\end{figure}

We present the results of the bipartite entanglement entropy for the subsystem $A$ here. The subsystem contains consecutive $L/2$ sites as shown in Fig.~\ref{fig:1D_model}(b). 
According to Eq.~(\ref{entropy}), the EE is closely related to the spectra of the correlation matrix. As shown in Fig.~(\ref{smImagScaling}), we find that the imaginary part of the spectra of the model described by Eq.~(\ref{Eq:H1d}) at the half-filling case is very small and vanish with the increase of the system size. Moreover, all the other cases discussed in this paper have the same behavior, which ensures that the EE is well-defined. These results indicate that these non-Hermitian systems can realize entanglement Hamiltonian with real eigenvalue spectra although they are governed by the non-unitary dynamics.

\begin{figure}[b]
	\includegraphics[angle=0,scale=0.099]{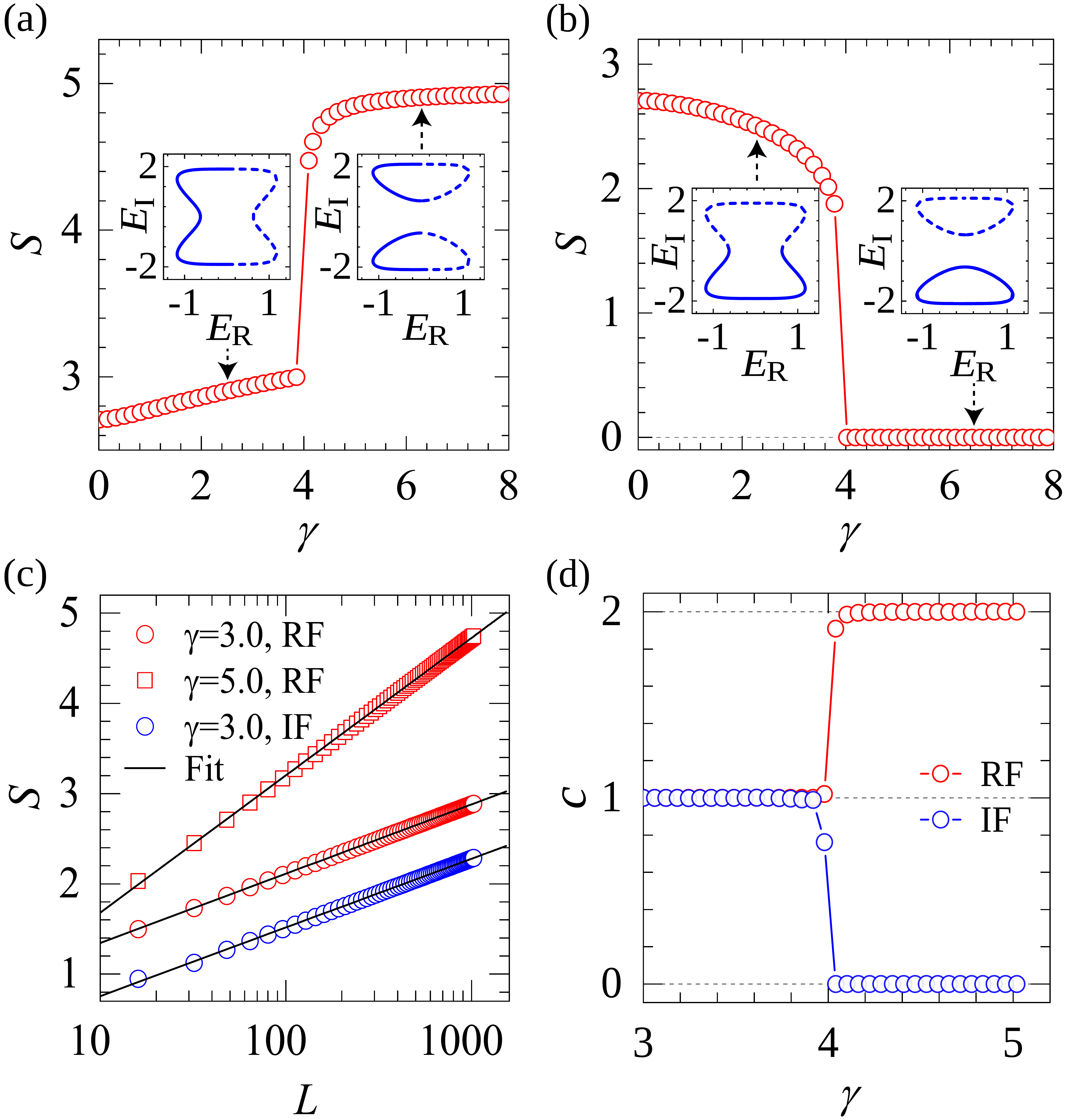}
	\caption{The entanglement entropy and $c$ for the $2$-sublattice model described by Eq.~(\ref{Eq:H1d}) at the half-filling case. The entanglement entropy as a function of $\gamma$ for the RF case (a) and IF case (b), where $L=1200$. The insets show the typical energy spectra, in which the solid (dashed) line denotes the occupied (unoccupied) states and $E_{R}$ ($E_{I}$) denotes the real (imaginary) part of the energy spectra. The energy spectra are demonstrated on both sides of the transition point. (c) The logarithmic fitting of the entanglement entropy. (d) $c$ as a function of $\gamma$ from the logarithmic fitting.}
	\label{fig:1d}
\end{figure}

When $n=2$, the model reduces to the well-known non-Hermitian Su-Schrieffer-Heeger (SSH) model\cite{SSH0a,SSH0b,klett2017relation,SSH3,SSH4,SSH5,SSH6}. At the half-filling case, the number of Fermi points of this SSH model will change as $\gamma$ increases, as shown in the inset of Fig.~\ref{fig:1d}(a) and Fig.~\ref{fig:1d}(b), thus the system undergoes a Lifshitz phase transition. Meanwhile, $S$ signals this phase transition and change dramatically across the critical point $\gamma_c$. The behaviors of the EE are different for the two kinds of definitions about the ground state. In the RF case, the EE increases significantly at $\gamma>\gamma_c$ since the number of Fermi points doubles. In the IF case, the EE becomes exact zero due to the point that the system becomes an insulator without any Fermi point. The above results reveal that the entanglement
entropy is only dependent on the geometry of the
Fermi surface and comes from the contributions of gapless modes.

Furthermore, we analyze the finite-size scaling of the EE as shown in Fig.~\ref{fig:1d}(c). Except for the insulator state, we find that the scaling behavior accurately satisfies the logarithmic function Eq.~(\ref{eq:Exact_SA}). Besides, $c$ is proportional to the number $N_f$ of Fermi points as shown in Fig.~\ref{fig:1d}(d), except near the transition point (the exceptional point), where strong singularity
obstructs the calculation of the EE. Thus the above results provide strong evidence that the asymptotic formula Eq.~(\ref{eq:Exact_SA}) is still valid for general 1D non-Hermitian free fermionic systems.

\begin{figure}[t]
\includegraphics[width=0.46\textwidth]{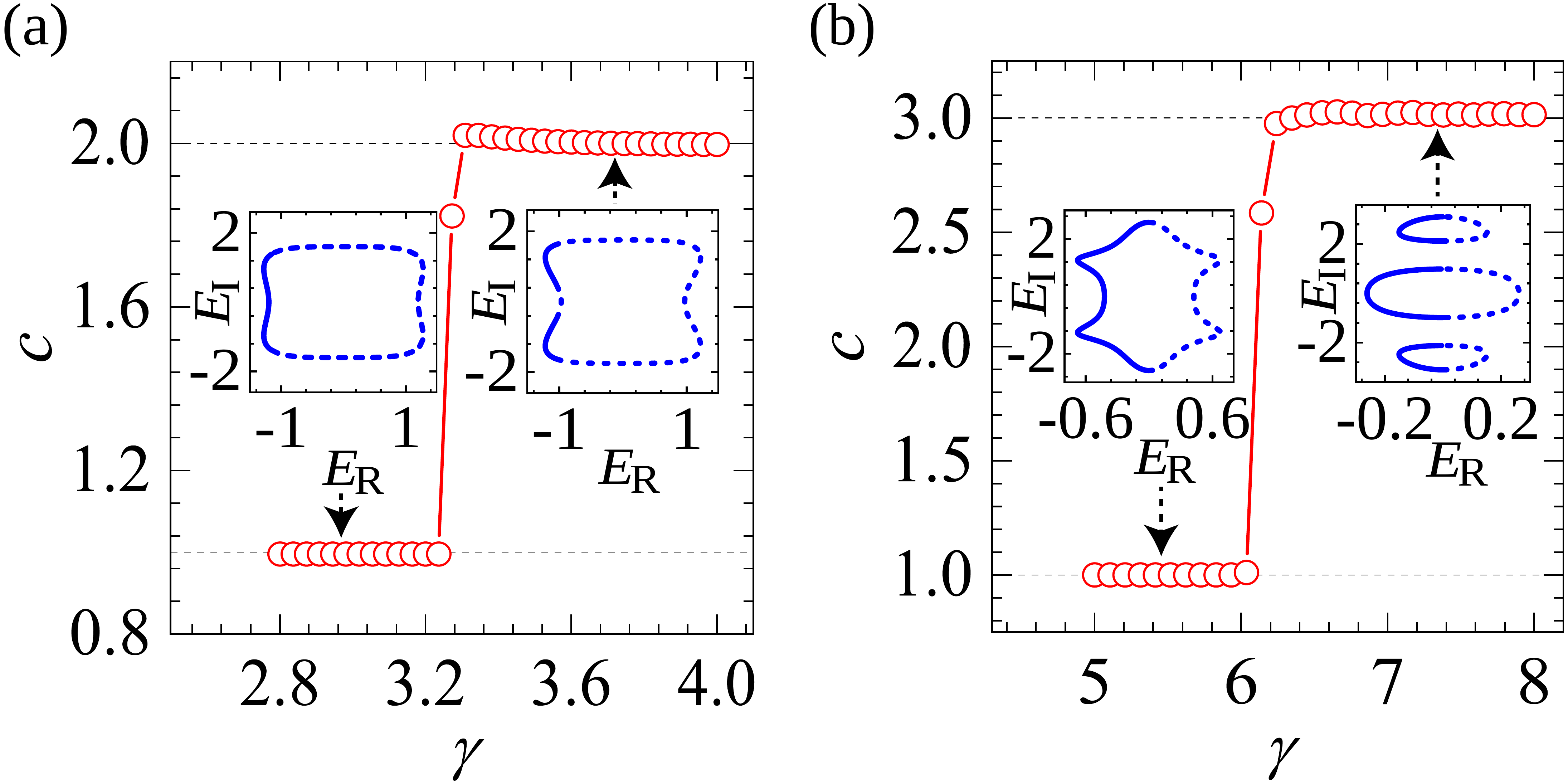}
\caption{(a) $c$ versus $\gamma$ for the $1/4$-filling $2$-sublattice model described by Eq.~(\ref{Eq:H1d}). (b) $c$ versus $\gamma$ for the half-filling $3$-sublattice model Eq.~(\ref{Eq:H1d}). The insets show the energy spectra, in which the solid (dashed) line denotes the occupied (unoccupied) states and $E_{R}$ ($E_{I}$) denotes the real (imaginary) part of the energy spectra. }
\label{fig:1d_cmpx}
\end{figure}

To confirm this point, we study the cases with different fillings of the model described by Eq.~(\ref{Eq:H1d}), which form a more complicated structure of the Fermi surface. Fig.~\ref{fig:1d_cmpx}(a) shows the results of $c$ change with $\gamma$ in the $1/4$-filling case, where the model is the Eq.~(\ref{Eq:H1d}) with $2$-sublattice. As shown in the inset of the Fig.~\ref{fig:1d_cmpx}(a), with $\gamma$ increases, the Fermi sea splits into two pieces, and the number of Fermi points doubles. As a result, one can observe that $c$ changes from $c=1$ to $c=2$ and keep the proportional relation on the number of Fermi points. 
For the half-filling case of the 3-sublattice model described by Eq.~(\ref{Eq:H1d}), the proportional relation is varified again, as shown in Fig.~\ref{fig:1d_cmpx}(b).

To conclude, the EE is proportional to the logarithmic of the system size, and $c$ is equal to half of the number of Fermi points.

\subsubsection{Eigenspectra of the Correlation Matrix}
\label{subsec:EigenOfCM}
\begin{figure}[b]
	\includegraphics[angle=0,scale=0.105]{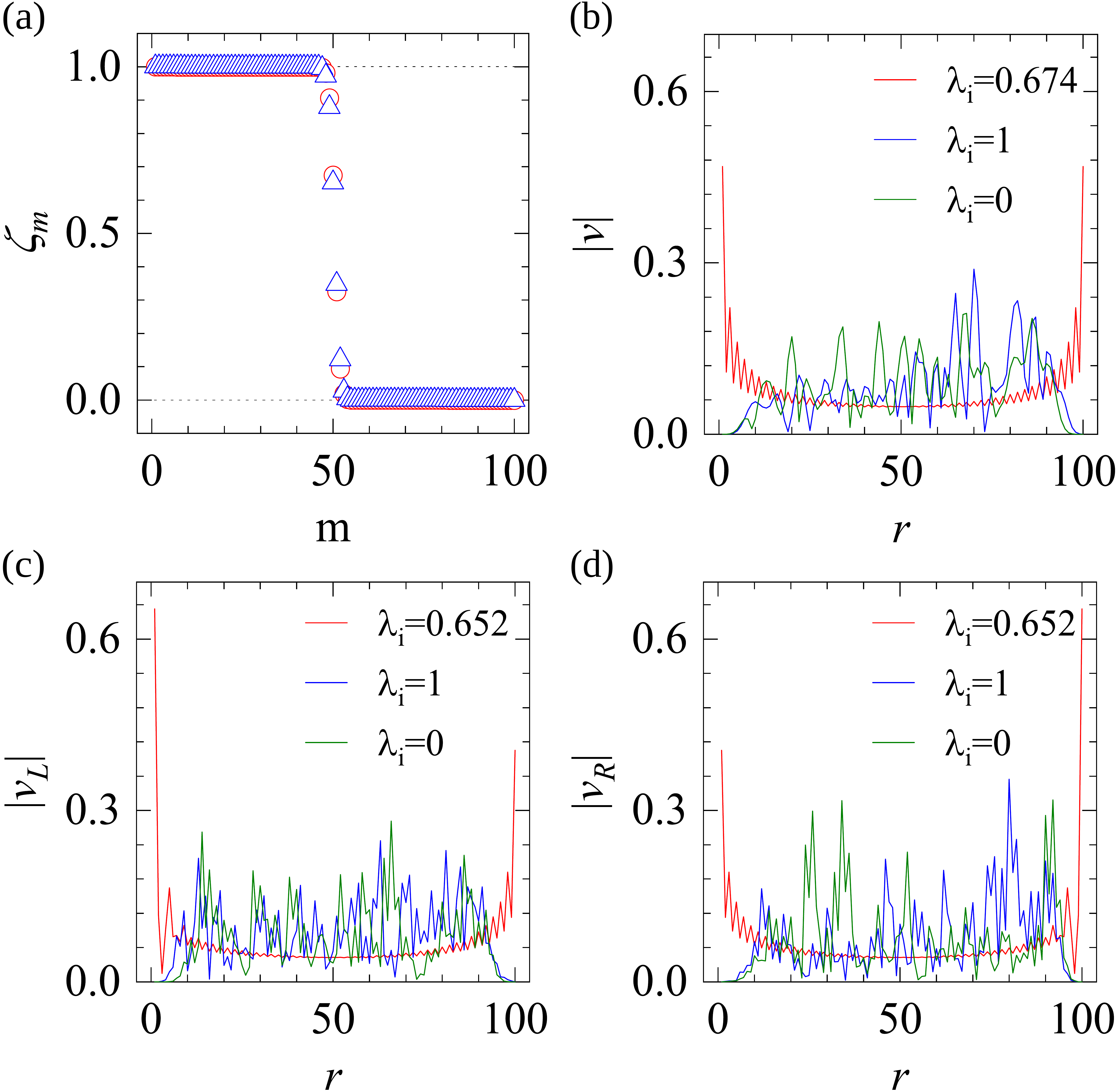}
	\caption{(a) The eigenspectra of the correlation matrix defined on subsystem $A$ in Hermitian(circle marker) and non-Hermitian(triangle marker) free fermionic system. (b) The real-space distribution of the eigenstate $|v|$ of the correlation matrix in the Hermitian case. (c-d) The real-space distribution of the left eigenstate ($|v_{L}|$) and right eigenstate ($|v_{R}|$) respectively in the non-Hermitian case. The Hermitian system in (a) and (b) is the free fermionic chain with the nearest neighbor hopping, and the non-Hermitian system in (c) and (d) is the half-filling $2$-sublattice model described by Eq.~({\ref{Eq:H1d}}).}
	\label{ESvector}
\end{figure}

To find the source of contribution to the EE, we analyze the general properties of eigenspectra $\zeta_i$ and the eigenstates $v_{L,R}$ of the correlation matrix $C^A$. Similar to the Hermitian fermion system, most $\zeta_i$ are close to $0$ or $1$, which do not contribute to the EE, as shown in Fig.~(\ref{ESvector})(a). We present the real-space distribution of the left and the right eigenstates of $C_A$ in Fig.~(\ref{ESvector})(c) and Fig.~(\ref{ESvector})(d). One can find that the typical eigenstates with $0$ or $1$ spectrum mainly distribute in the bulk of $A$. They decrease dramatically when approaching the boundary of subregion $A$. The typical eigenstates which make the main contribution to the EE, locate predominately near the boundary. It is consistent with the conclusion from the Hermitian case, where the eigenstate which contributes to the EE carries the entangled pairs across the boundary\cite{fishman2015compression}. Due to the entangled pairs, the related eigenstates must have a large magnitude near the boundary. While the entangled pairs are absent in the eigenstates with spectrum $0$ or $1$, leading to the main distribution in the bulk. For comparison, we present the eigenstates of the Hermitian systems in Fig.~(\ref{ESvector})(b).

\subsection{$2D$ System}
We continue to discuss the entanglement properties of two-dimensional ($2D$) non-Hermitian free fermionic lattice models. The $2D$ model is a generalization of the above $1D$ model. This $2D$ Hamiltonian on the square lattice reads
\begin{align}
H_{2D}
= &\, \sum_{i, j} \big[ (c_{2i,j}^{\dagger}c_{2i+1,j}
+ c_{j,2i}^{\dagger} c_{j,2i+1} + h.c.)  \nonumber \\
+ & \,\,  t^L ( c_{2i-1,j}^{\dagger}c_{2i,j} + c_{j,2i-1}^{\dagger} c_{j,2i}) \nonumber \\
+ & \,\,  t^R (c_{2i,j}^{\dagger} c_{2i-1,j} + c_{j,2i}^{\dagger} c_{j,2i-1}) \big],
 \label{Eq:H2d}
\end{align}
where $t^L = 1 + \frac{\gamma}{2}$ and $t^R = 1 - \frac{\gamma}{2}$. 

\begin{figure}[b]
\includegraphics[width=0.47\textwidth]{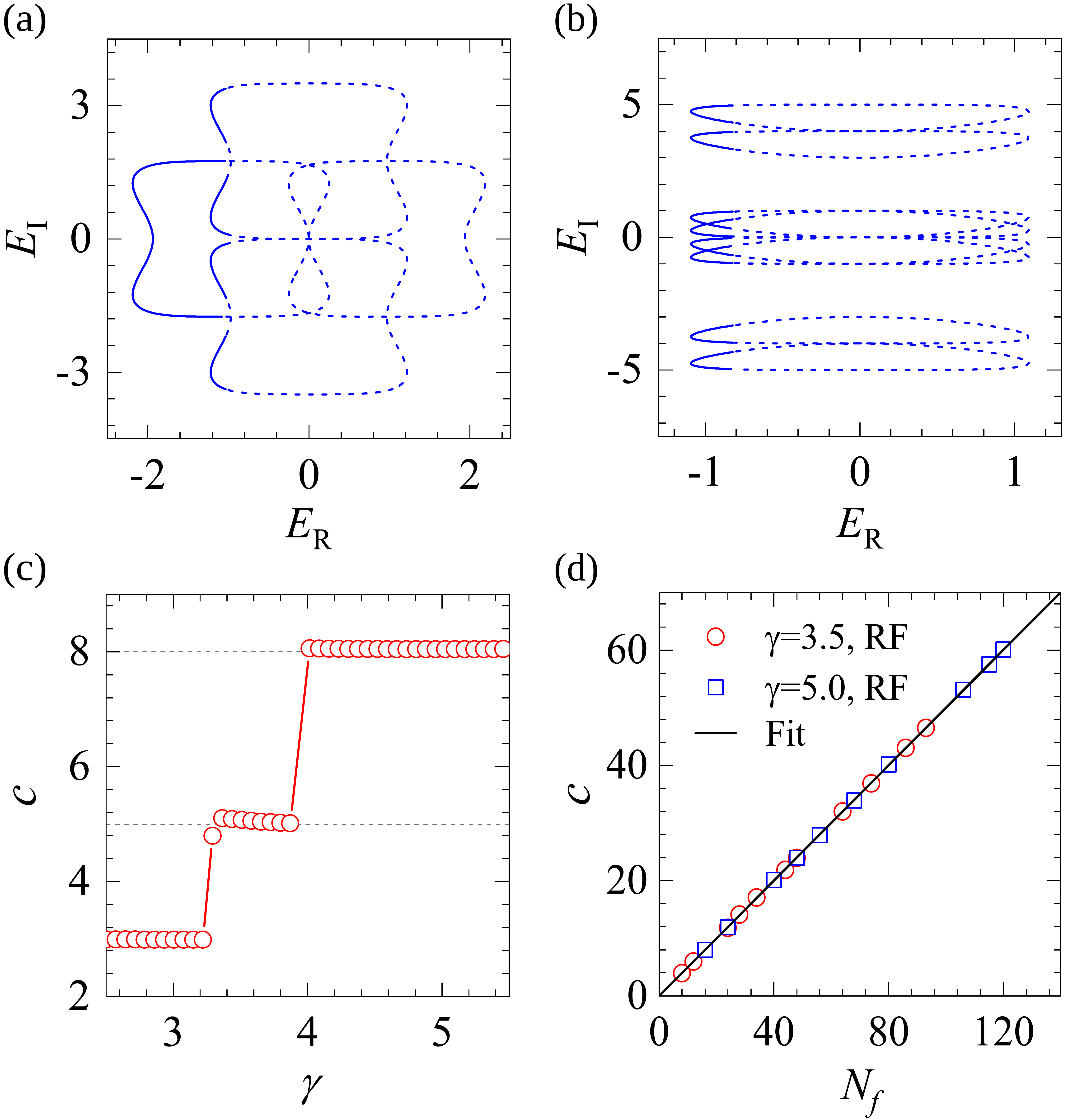}
\caption{The entanglement entropy for the ladder described by Eq.~(\ref{Eq:H2d}). The energy spectra for the $1/4$ filling $L_y=4$ ladder at $\gamma=3.5$ (a) and $\gamma=5.0$ (b), in which the solid (dashed) line denotes the occupied (unoccupied) states and $E_{R}$ ($E_{I}$) denotes the real (imaginary) part of the energy spectra. (c) $c$ from the logarithmic fitting for the $1/4$-filling $L_y=4$ ladder as a function of $\gamma$. (d) $c$ from the logarithmic fitting as a function of the number of Fermi points $N_f$ for the half-filling case. There are approximate relations between $N_f$ and $L_y$: $N_f \approx 3 L_y$($N_f \approx 4L_y$) at $\gamma=3.5$($\gamma=5.0$) when $L_y$ is large.}
\label{fig:ladder}
\end{figure}

Before focusing on the true $2D$ case, we discuss the following quasi-one-dimensional case: the $L_y=4$ ladder with $1/4$-filling.
As shown in Fig.~\ref{fig:ladder}(a) and Fig.~\ref{fig:ladder}(b), this model has more complicated structures of energy spectra. The number of Fermi points changes from $N_f=6$ to $N_f=10$ then $N_f=16$ with the increase of $\gamma$. Besides, from the logarithmic fitting of the $S$ with $L_x$, it is shown again that $c$ is equal to half of the number of Fermi points, sees in Fig.~\ref{fig:ladder}(c). We further study the behavior of $c$ with the increase of width $L_y$ in the half-filling case. In Fig.~\ref{fig:ladder}(d), it is observed that $c$ is proportional to the number of Fermi points $N_f$ regardless of the parameter $\gamma=3.5$ or $5.0$. These results again confirm that the finite-size scaling behavior of the EE obeys the logarithmic form in Eq.~(\ref{eq:Exact_SA}) and $c$ is equal to $N_f/2$, irrelevant to the detailed energy spectra structure.
 
\begin{figure}[t]
\includegraphics[width=0.475\textwidth]{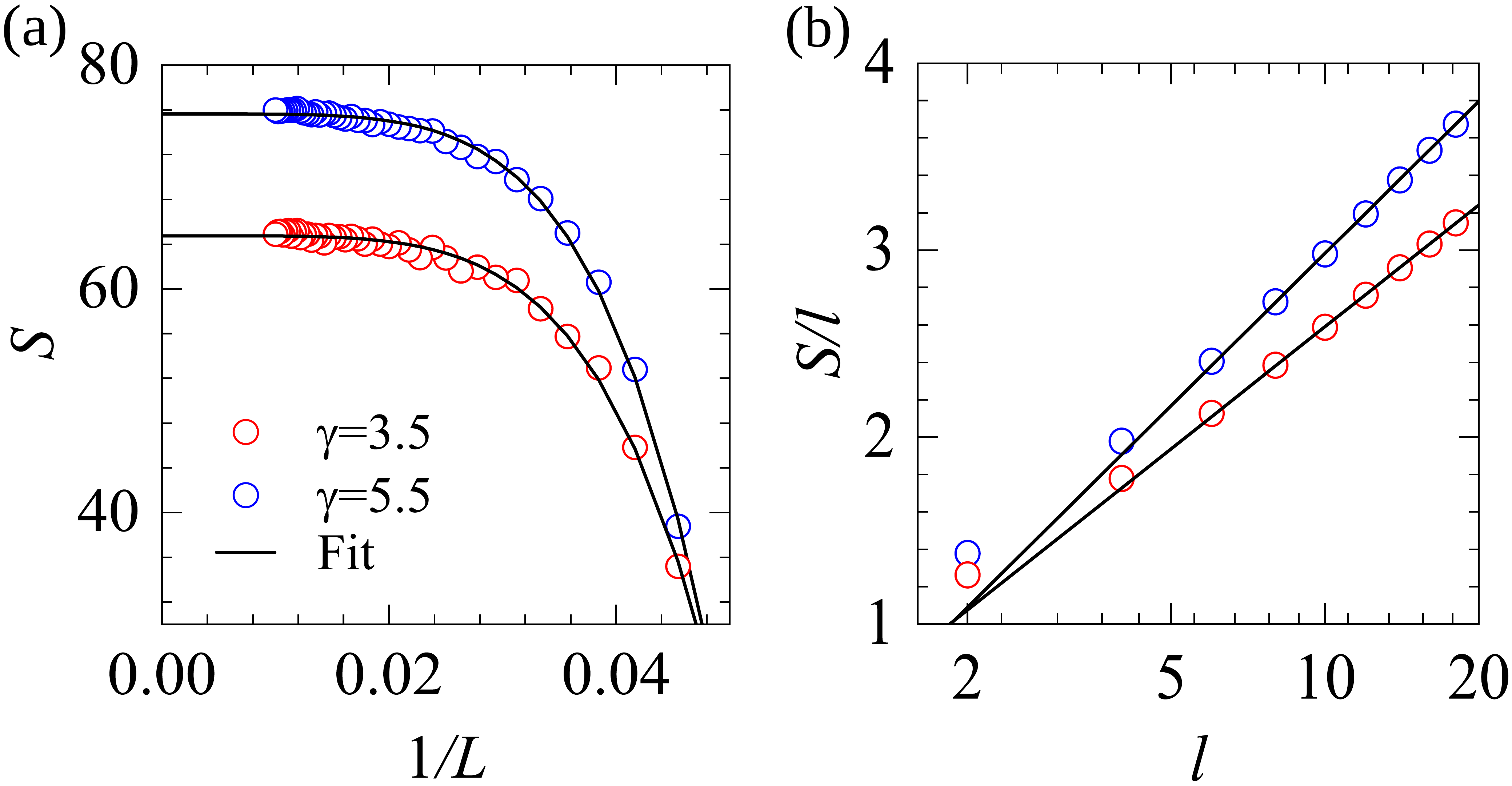}
\caption{The entanglement entropy of the $2D$ model defined by Eq.~(\ref{Eq:H2d}) for the real half-filling ground state. (a) The extrapolation of the entanglement entropy for a given $l \times l$ subregion with $l=20$. (b) The logarithmic fitting for $S/l$ to the subsystem size $l$.}
\label{fig:2d}
\end{figure}

Finally, we study the EE between a $l \times l$ subsystem $A$ and the other part $\bar{A}$ of the square lattice as shown in Fig.~\ref{fig:1D_model}(d). It is expected that the entanglement entropy should depend on both the subsystem size $l$ and the total system size $L$. 
We first analyze the scaling behavior of the total system size $L$ for a given subsystem size $l$. Then from the extrapolation, we obtain $S(l)$ as shown in Fig.~\ref{fig:2d}(a).
Eventually, we find that the EE $S(l)/l$ increases linearly with $\log(l)$ as shown in Fig.~\ref{fig:2d}(b), namely 
\begin{equation}
S \sim l \, \mathrm{log}(l).
\label{Eq:S2d}
\end{equation}
Eq.~(\ref{Eq:S2d}) indicates that the EE has an additional logarithmic term to the area law in the $2D$ non-Hermitian model described by Eq.~(\ref{Eq:H2d}). The behavior is explained by the fact that the characteristic length here is $l$. This result is similar to that of the $2D$ Hermitian free fermionic system with a finite Fermi surface\cite{swingle2010entanglement}.

\subsection{Correlation functions}
\begin{figure}[tbp]
	\includegraphics[angle=0,scale=0.09]{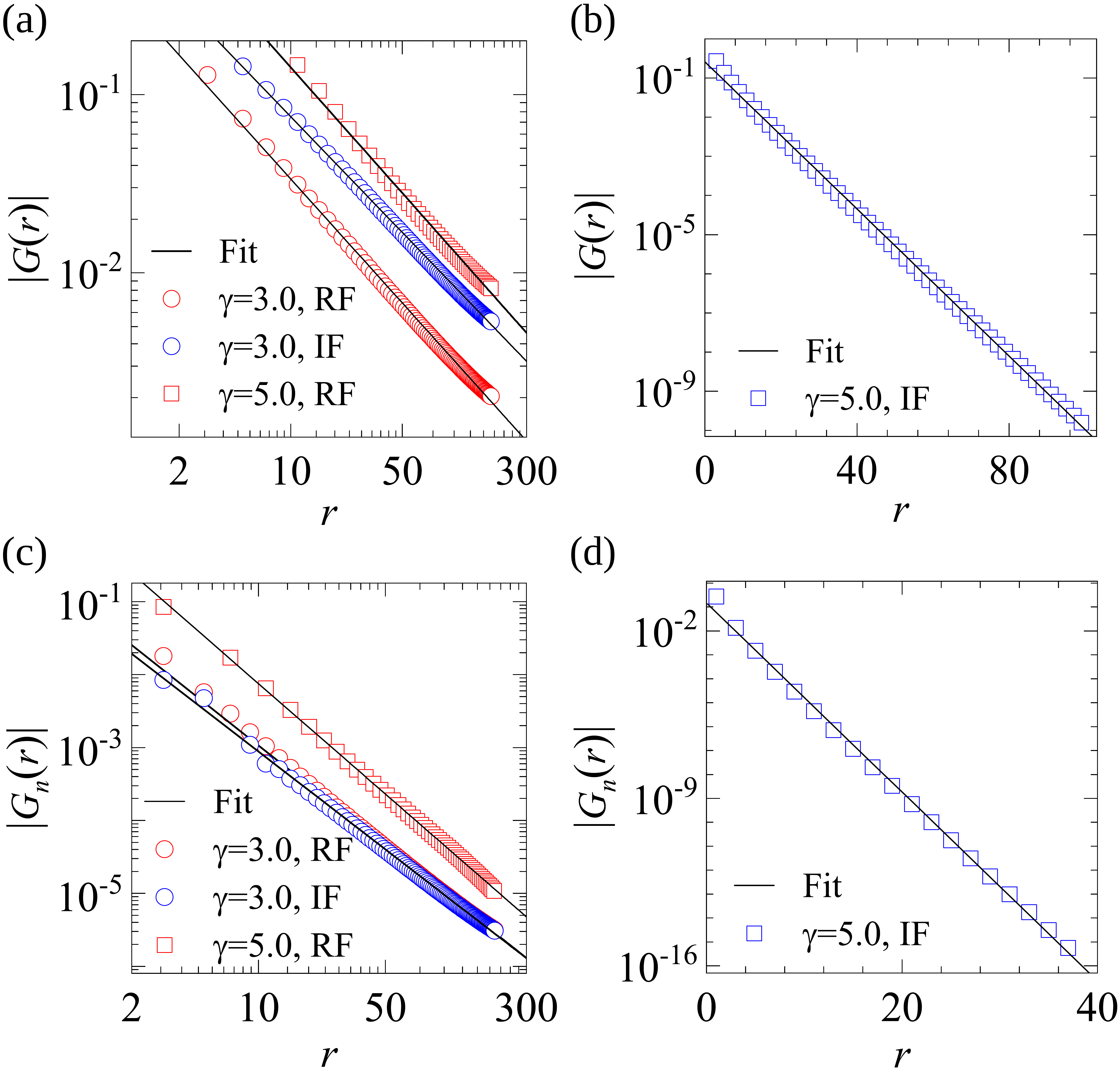}
	\caption{The correlation functions for the $2$-sublattice model defined by Eq.~(\ref{Eq:H1d}) at half-filling. (a) Power-law and (b) exponential fitting for the single-particle correlation function. (c) Power-law and (d) exponential fitting for the density-density correlation function.}
\label{smfig:correlation}
\end{figure}

In Hermitian free fermion systems, there are close connections between the entanglement and the correlations of scaling operators\cite{amico2008entanglement,brandao2013area,hastings2007entropy}. The important entanglement information of low energy physics is encoded in the correlations. Thus, it is important to explore the correlations in the non-Hermitian systems. Specially, we study the single-particle and the density-density correlation function of the metallic states with Fermi points and insulating states without Fermi points.
The single-particle correlation function $G(r_{ij})$ is nothing but the correlation matrix $C^{A}_{ij}$, and the density-density correlation function is expressed as 
\begin{equation}
G_{n}(r) = \langle G_L \vert n_1 \, n_{r+1} \vert G_R \rangle - \langle G_L \vert n_1 \vert G_R \rangle \langle G_L \vert n_{r+1} \vert G_R \rangle,
\end{equation}
where $\langle G_L|$ and $|G_R \rangle$ indicate the left and the right ground state. According to Wick's theorem, the density-density correlation functions can be decomposed into two single-particle correlation functions $G_n(r) = \langle G_L \vert c_1^\dagger c_{r+1} \vert G_R \rangle \langle G_L \vert c_1 c_{r+1}^\dagger \vert G_R \rangle$. As results, the correlation function of a metallic state shows a power-law behavior, whereas the correlation function of an insulating state decays exponentially, as shown in Fig.~\ref{smfig:correlation}. In Table~\ref{tablel1}, we present the scaling dimensions $\Delta$ and $\Delta_n$ of the fermion operator and the density operator, respectively, from the following power-law fittings
\begin{equation}
\left \vert G(r) \right \vert \sim r^{-2 \Delta}, \quad \left \vert G_n(r) \right \vert \sim r^{-2 \Delta_n}.
\end{equation}
Note that $\Delta$ should be regarded as the average of the scaling dimensions of the $c_L$ and $c_R^\dagger$ operators since they are not conjugate to each other. Interestingly, the scaling dimensions evaluated from the power-law fitting are approximate to be $\Delta=1/2$ and $\Delta_n=1$, which are the same as that of the Hermitian free fermions.

\renewcommand\arraystretch{1.4}
\begin{table}
\begin{center}
\caption{The scaling dimension of the fermion ($\Delta$) and density operator ($\Delta_n$)}
\setlength{\tabcolsep}{1.55mm}
\begin{tabular}{p{1cm}<{\centering}|p{2cm}<{\centering}|p{2cm}<{\centering}|p{2cm}<{\centering}}
\hline
\hline
$\quad$ & $\gamma=3.0$, RF & $\gamma=5.0$, RF & $\gamma=3.0$, IF\\
\hline
${\Delta}$ & 0.50 & 0.50 & 0.46\\
\ ${\Delta_n}$ & 0.99 & 1.08 & 0.96\\
\hline
\hline
\end{tabular}
\label{tablel1}
\end{center}
\end{table}

\section{Summary}
\label{sec:summary}
We have investigated the entanglement properties of the non-Hermitian free fermionic models in one and two dimensions. For the one-band model, we prove that the EE increases logarithmically with the system size $L$ and each gapless mode contributes $\frac{1}{2}$ to the coefficient $c$ of the logarithmic correction. These results are numerically confirmed for more general non-Hermitian free fermionic systems with more complicated energy spectra in both one and two dimensions. 
They hold for both the RF and IF types of ground states, which reveals that the entanglement entropy is only dependent on the geometry of the Fermi surface.
Analyzing the spectra and eigenstates of the subsystem correlation matrix, we illustrate that the EE comes from the entangled pairs across the boundary.
In addition, we extract the scaling dimensions of fermion and density operators by the correlation functions.

Our work could promote the understanding of the entanglement properties in general non-Hermitian systems. 
On the one hand, it provides a reference to explore the entanglement properties in interacting non-Hermitian systems. It would be very intriguing to explore whether these properties found in this paper still hold when the interaction is included. On the other hand, the entanglement results here are very similar to that of the Hermitian case, which may serve as an inspiring example to seek the potential deep connections between non-Hermitian and Hermitian systems.
Also, there exists the possibility of experimental realization, and it’s worth trying to realize our models in the existing techniques\cite{zhang2018parity,zhong2017transport,li2017controlled}.
Still, there are some issues to be resolved. The ground states of these models somehow possess real entanglement spectra. It is natural to pursue what the restrictions are to guarantee the real entanglement spectra for the non-Hermitian free fermions. Another follow-up question is how to define the EE when the complex entanglement spectra occur. We leave these open questions for future study.

\section{Acknowledgments}
We thank Fan Cui, Guang-Lei Xu, Xing-Jie Han and
Long Zhang for helpful discussions.
The work is supported by the National Key Research and Development Project of China (Grant No. 2017YFA0302901), the National Natural Science Foundation of China (Grants No.~11888101 and No.~11874095), the Youth Innovation Promotion Association CAS (Grants No.~2021004), and the Strategic Priority Research Program of Chinese Academy of Sciences (Grant No. XDB33000000). R.Z.H is supported by China Postdoctoral Science Foundation (Grant No. 2020T130643), the Fundamental Research Funds for the Central Universities and the National Natural Science Foundation of China (Grants No. 12047554). Y.C.Y is supported by the National Science Foundation (NSF) for Young Scientists of China under Grant No.11804377. L.P.Yang is supported by the Natural Science Foundation of Chongqing (Grant No. cstc2018jcyjAX0399).

\bibliography{main}

\appendix
\twocolumngrid
\section{Non-Hermitian free fermions with translational symmetry}
\label{subsec:ExactDiag}
For translational invariant free fermionic models with periodic boundary conditions,
the Hamiltonian can be transformed to block matrix via unitary transformations. For each momentum $k$, the Hamiltonian of the $n$-sublattice models defined in (\ref{Eq:H1d}) should be a $n\times n$ matrix. One may transform the origin fermion operators to the new fermion operators $\tilde{c}_{m}$ by unitary transformation
\begin{equation}
c_j={\sum}_{m} F_{jm} \tilde{c}_{m}.
\end{equation}
Here $m=1,...,N-1,N$ and the unitary transformation
\begin{equation}
F_{jm} = N^{-1/2}\exp\left[\frac{-2\pi\mathrm{i}(m-1)(j-1)}{N}\right].
\end{equation}

Take the $2$-sublattice model Eq.~({\ref{Eq:H1d}}) for example, the block matrix $H^{\prime} = F^{\dagger}HF$ is read as
\begin{equation}
H^{\prime} = 
\left(
\begin{array}{cc}
H_{A}^{^{\prime }} & -H_{B}^{^{\prime }} \\
H_{B}^{^{\prime }} & -H_{A}^{^{\prime }}%
\end{array}%
\right),  \label{Block_Hamiltonian}
\end{equation}
where 
\begin{eqnarray}
\left[H^{\prime}_{A}\right]_{p,q} &=& \delta_{p,q}A(k_q), \notag \\
\left[H^{\prime}_{B}\right]_{p,q} &=& \delta_{p,q}B(k_q)
\end{eqnarray}
and
\begin{eqnarray}
A(k_q) &=& 2\cos(k_q) - \mathrm{i}\frac{\gamma}{2}\sin(k_q),  \notag \\
B(k_q) &=& \frac{\gamma }{2}\cos(k_q)
\label{AB}
\end{eqnarray}
with $k_q = \frac{2\pi(q-1)}{N}$.

Then the spectra are directly obtained by diagonalizing (\ref{Block_Hamiltonian}). To be compact, the spectra, the corresponding right, and left eigenvectors are arranged in matrix $\mathcal{D},$$ L, R$ such that $L\mathcal{D}R = H^\prime$. The spectra
\begin{equation}
\mathcal{D} = 
\left(
\begin{array}{cc}
-D & 0 \\
0 & D%
\end{array}%
\right),  \label{spectrum}
\end{equation}
with
\begin{eqnarray}
    D_{pq} &=& \delta_{p,q} \sqrt{A(k_q)^{2}-B(k_q)^{2}}
\end{eqnarray}
is denoted as
\begin{equation}
    D(k_{q}) = D_{qq}.
\end{equation}
The corresponding right eigenvectors are written as
\begin{equation}
L = 
\left(
\begin{array}{cc}
L^{--} & L^{-+} \\
L^{+-} & L^{++}%
\end{array}%
\right)
\end{equation}
with
\begin{eqnarray}
    \left[L^{-\pm}\right]_{p,q} &=& \delta_{p,q} n_{q}\left[A(k_q)\pm D(k_q)\right], \notag \\
    \left[L^{+\pm}\right]_{p,q} &=& \delta_{p,q} n_{q}B(k_q).
\end{eqnarray}
The left eigenvectors are expressed as
\begin{equation}
R = 
\left(
\begin{array}{cc}
R^{--} & R^{-+} \\
R^{+-} & R^{++}%
\end{array}%
\right)
\end{equation}
with
\begin{eqnarray}
    \left[R^{\pm-}\right]_{p,q} &=& \pm\delta_{p,q} n_{q} B(k_{q}), \notag \\
    \left[R^{\pm+}\right]_{p,q} &=& \delta_{p,q}n_{q} \left(D(k_q) \mp A(k_{q}) \right).
\end{eqnarray}
Where $p,q=1,2,..,N/2$ and normalized factor $n_q=[2B(k_q)D(k_q)]^{-1/2}$. Here, $\delta_{pq}$ is the Kronecker delta function.

Then the correlation matrix can be represented by the matrix
$L, R$, and $F$ and reads
\begin{equation}
C^{A} _{mn}=\sum\limits_{\substack{ 1\leq p,s \leq N   \\ q\in \mathcal{S}}} F_{np}L_{pq}R_{qs}F^{\dagger}_{sm}.
\label{cor}
\end{equation}%
It will become extremely challenging to analytically calculate the EE for the $2$-sublattice model defined by Eq.~(\ref{Eq:H1d}). The reason is that the correlation matrix here turns out to be a block-Toeplitz matrix in the continuous limit~\cite{R2010A, Its2006Entropy, Steve2000Toeplitz}. For the $2\times 2$ block that locates at the i-th row and j-th column in the correlation matrix, we have
\begin{align}
C^{A}_{ij}=\frac{1}{2\pi }\int_{0}^{\pi }\mathrm{d}p \,
\mathrm{e}^{\mathrm{i} 2p(i-j)}T(p)
\label{Cmn}
\end{align}
with $T(p)$ defined by
\begin{equation}
T(p)=\left(
\begin{array}{cc}
1 & -e^{-\mathrm{i}p}/t(p) \\
-e^{\mathrm{i}p}t(p) & 1
\end{array}
\right)
\label{Tq}
\end{equation}
and
\begin{equation}
t(p) = \frac{A(p)+B(p)}{D(p)}
\end{equation}
Notice that the symbol of this block-Toeplitz matrix is not homomorphically defined on the whole Riemann surface~\cite{Its2006Entropy}. However, we can take a detour to calculate the entanglement entropy by numerical calculation. 

In general, the correlation matrix of other models such as general $n$-sublattice models defined by Eq.(\ref{Eq:H1d}) and two-dimensional models defined by Eq.(\ref{Eq:H2d}) can be derived similarly.
As expected, one would still encounter the same fatal problem in the $2$-sublattice model here during the analytical deduction. But the numerical simulation for
the entanglement entropy from the correlation matrix $C^A$ is still feasible. 

\end{document}